\documentclass[12pt]{article}
\usepackage{amsmath}
\usepackage{amssymb}
\usepackage[mathscr]{eucal}
\usepackage[usenames]{color}
\usepackage[latin5]{inputenc}
\usepackage{url}
\definecolor{webgreen}{rgb}{0,0.5,0}

\newtheorem{thm}{Theorem}

\newtheorem{rmk}{Remark}

\numberwithin{equation}{section}
\numberwithin{thm}{section}
\numberwithin{lemma}{section}
\numberwithin{prop}{section}
\numberwithin{cor}{section}
\numberwithin{rmk}{section}
\numberwithin{defn}{section}

\newcommand{\curl}[1]{ \{#1\} }

\definecolor{darkolivegreen}{rgb}{0.333333, 0.419608, 0.1843140}

\DeclareMathOperator{\Sl}{sl} 
 
\DeclareMathOperator{\SL}{SL}

\newcommand{\dx}{\partial_x}
\newcommand{\dy}{\partial_y}
\newcommand{\dt}{\partial_t}
\newcommand{\du}{\partial_u}
\newcommand{\dv}{\partial_v}
\newcommand{\dw}{\partial_w}

\newcommand{\der}{\frac{d}{dt}}

\begin{document}
\pagenumbering{arabic}

\title{\Large Variable coefficient Davey-Stewartson  system with a Kac-Moody-Virasoro symmetry algebra}

\author{
F.~G\"ung\"{o}r\\ \small
Department of Mathematics, Faculty of Science and Letters,\\ \small Istanbul Technical University, 34469 Istanbul, Turkey \thanks{e-mail: gungorf@itu.edu.tr}\and
C. \"{O}zemir\\
\small Department of Mathematics, Faculty of Science and Letters,\\
\small Istanbul Technical University, 34469 Istanbul,
Turkey \thanks{e-mail: ozemir@itu.edu.tr} }

\date{\today}
\clearpage
\maketitle
\thispagestyle{empty}

\begin{abstract}
We study the symmetry group properties of the variable coefficient Davey-Stewartson (vcDS) system. The Lie point symmetry algebra with a Kac-Moody-Virasoro (KMV) structure is shown to be isomorphic to that of the usual (constant coefficient) DS system if and only if the coefficients satisfy some conditions. These conditions turn out to coincide with those for the vcDS system to be transformable to the DS system  by a point transformation.

The equivalence group of the vcDS system is  applied to pick out the integrable subsystems from a class of non-integrable ones. Additionally,  the full symmetry group of the DS system  is derived explicitly without exponentiating its symmetry algebra.

Lump solutions (rationally localized in all directions in  $\mathbb{R}^2$) introduced by Ozawa for the DS system is shown to hold even for the vcDS system precisely when the system belongs to the  integrable class, i.e. equivalent to the DS system. These solutions can be used for establishing exact blow-up solutions in finite time in the space $L^2(\mathbb{R}^2)$ in the focusing case.
\end{abstract}

\section{Introduction}
The Davey-Stewartson (DS) system
\begin{subequations}\label{DS}
\begin{eqnarray}
   \label{DSa}  &&i\psi_t+\psi_{xx}+\epsilon_1\psi_{yy}=\epsilon_2|\psi|^2\psi+\psi w, \\
   \label{DSb}  &&w_{xx}+\delta_1w_{yy}=\delta_2(|\psi|^2)_{yy},
\end{eqnarray}
\end{subequations}
where $\epsilon_1=\pm 1$, $\epsilon_2=\pm 1$, $\delta_1,\delta_2$ are constants and $\psi$ is a complex and $w$ a real function of $t,x,y$, first appeared in the context of water waves \cite{DaveyStewartson1974}. A brief rederivation of \eqref{DS} can also be found in \cite{GhidagliaSaut1990}.

The Lie point symmetry algebra of the DS system
is known to be infinite-di\-men\-sion\-al with Kac-Moody-Virasoro structure only in the integrable case $\delta_1=-\epsilon_1$, namely, either in elliptic-hyperbolic (DS I) or hyperbolic-elliptic (DS II) case \cite{ChampagneWinternitz1988, Omote1988}. In the non-integrable case the Virasoro part is absent in the entire symmetry algebra. Painlev\'e analysis and Hirota's bilinearization of the integrable DS system was performed in \cite{GanesanLakshmanan1987}. The complete integrability  also implies the existence of infinite number of conservation laws and generalized symmetries. Some conservation laws (conservation of energy, mass and momentum) and a virial (or variance) identity, which plays a key role in proving existence of blow-ups in finite time $0<T<\infty$ for the focusing DS II system with a negative energy $E(u)=E(u_0)<0$ in the virial space $u\in H^1\cap\curl{(x^2+y^2)^{1/2}|u|\in L^2}$, $t\in [0,T)$,  were derived in \cite{GhidagliaSaut1990}.

In this article we are concerned with a study of Lie symmetry properties of a generalization of \eqref{DS} obtained by including arbitrary time dependent coefficients.   We consider  variable coefficient Davey-Stewartson (vcDS) equations
in the form
\begin{subequations}\label{VCDS}
\begin{eqnarray}
   \label{VCDSa}  &&i\psi_t+p_1(t)\psi_{xx}+p_2(t)\psi_{yy}=q_1(t)|\psi|^2\psi+q_2(t)\psi w, \\
   \label{VCDSb}  &&w_{xx}+r_1(t)w_{yy}=r_2(t)(|\psi|^2)_{yy},
\end{eqnarray}
\end{subequations}
where all the coefficients are real non-zero arbitrary  smooth functions of $t$.  We wish to determine the coefficients under which \eqref{VCDS} is invariant under  an infinite-dimensional symmetry algebra which is isomorphic to that of \eqref{DS}. We also use the equivalence group of \eqref{VCDS} to determine the subclasses  when the vcDS system is locally equivalent to its constant coefficient counterpart. We establish that there is a close link between the existence of the KMV symmetry algebra structure  and a point transformation taking \eqref{VCDS} to the integrable DS system. It is also known that the 3-wave form of \eqref{DS} which arose in a different context admits an isomorphic symmetry algebra \cite{GuengoerAykanat2006, LiYeChen2008}.

We note that  a study of bilinear form, Baecklund transformation and Lax pair for \eqref{VCDS} was given in Ref. \cite{ZhouTianMoLiWang2013} under some restriction of the coefficients. We do not find this restriction in agreement with any of the conditions in our framework.

The organization of the paper is as follows. In Section 2 the equivalence group  $\mathsf{G_E}$ of \eqref{VCDS} is constructed.  Transformation to some canonical form is achieved by use of $\mathsf{G_E}$. A discussion of the derivation of the symmetry group for the DS system from $\mathsf{G_E}$ is also given. Section 3 is devoted to an analysis of Lie symmetry properties of \eqref{VCDS}. Section 4 discusses some solutions based on equivalence and symmetry groups. In particular, it is shown that how Ozawa's exact lump solutions \cite{Ozawa1992}  can be used in this context.  Finally, the results obtained are summarized in Section 5.

\section{Equivalence transformations}\label{section2}
We wish to determine the conditions on the coefficients for the system \eqref{VCDS} to be locally equivalent to its constant coefficient counterpart \eqref{DS} under local point transformations. We will take advantage of the notion of equivalence transformations (or allowed transformations). We refer to \cite{GungorOzemir2015} for a short summary of these transformations. They are defined to be locally invertible transformations preserving the form of the equations \eqref{VCDS}. Here we restrict to only fiber-preserving transformations
\begin{equation}\label{fiber-preserv}
\tilde{t}=T(t, x, y),
\quad \tilde{x}=X(t, x, y),
\quad \tilde{y}=Y(t, x, y),
\quad \psi=\Psi(t,x,y,\tilde{\psi}), \quad w=W(t,x,y,\tilde{w}),
\end{equation}
where $\tilde{\psi}$ and $\tilde{w}$ are functions of the new coordinates $(\tilde{t},\tilde{x},\tilde{y})$. The transformations \eqref{fiber-preserv} will transform \eqref{VCDS} to
\begin{subequations}\label{VCDSnew}
\begin{eqnarray}
   \label{VCDSnewa}  &&i\tilde{\psi}_{\tilde{t}}+\tilde{p}_1(\tilde{t})\tilde{\psi}_{xx}+\tilde{p}_2(\tilde{t})\tilde{\psi}_{\tilde{y}\tilde{y}}=
   \tilde{q}_1(\tilde{t})|\tilde{\psi}|^2\tilde{\psi}+\tilde{q}_2(\tilde{t})\tilde{\psi} \tilde{w}, \\
   \label{VCDSnewb}  &&\tilde{w}_{\tilde{x}\tilde{x}}+\tilde{r}_1(\tilde{t})w_{\tilde{y}\tilde{y}}=\tilde{r}_2\tilde{(t})(|\tilde{\psi}|^2)_{\tilde{y}\tilde{y}}.
\end{eqnarray}
\end{subequations}
Requiring that the derivatives $\tilde{\psi}_{\tilde{t}}$, $\tilde{\psi}_{\tilde{x}\tilde{x}}$ and $\tilde{\psi}_{\tilde{y}\tilde{y}}$ be preserved linearly we obtain the linearity of $\Psi$ in $\tilde{\psi}$ and $W$ in $\tilde{w}$. The requirements that there are no terms $\tilde{t}_{tx}$ and $\tilde{t}_{ty}$ and the nonlinearity of $\psi$ be preserved imply that our equivalence transformations must have the form
\begin{equation}\label{equiv}
\begin{split}
&\tilde{t}=T(t), \quad \tilde{x}=X(t, x, y), \quad \tilde{y}=Y(t, x, y), \\[.5mm]
&    \psi=Q(t,x,y)\tilde{\psi},  \quad w=L(t,x,y)\tilde{w}+H(t,x,y),
\end{split}
\end{equation}
where $Q$ is a complex-valued, $L$ and $H$ are real-valued functions. We also assume that $\dot{T}X_xY_y\ne 0$ and $Q\ne 0$, $L\ne 0$ for invertibility. The transformations \eqref{equiv} will keep the differential form of the system intact if we further impose the following conditions (coming from the necessity that the expressions proportional to $\tilde{\psi}_{\tilde{x}}$, $\tilde{\psi}_{\tilde{y}}$ and $\tilde{\psi}$ must vanish and no other new terms added to the system)
\begin{equation}\label{constr123}
\begin{split}
&i Q X_t+p_1(Q X_{xx}+2Q_xX_x)=0, \quad i Q Y_t+p_1(Q Y_{yy}+2Q_yY_y)=0,\\[.5mm]
&iQ_t+p_1Q_{xx}+p_2Q_{yy}=q_2QH, \quad H_{xx}+r_1H_{yy}=0.
\end{split}
\end{equation}
These transformations change the coefficients as follows:
\begin{equation}\label{newcoeff}
    \tilde{p}_1=\frac{X_x^2}{\dot{T}}p_1, \quad \tilde{p}_2=\frac{Y_y^2}{\dot{T}}p_2, \quad \tilde{q}_1=\frac{|Q|^2}{\dot{T}}q_1, \quad \tilde{q}_2=\frac{L}{\dot{T}}q_2, \quad \tilde{r}_1=\frac{Y_y^2}{X_x^2}r_1,  \quad \tilde{r}_2=\frac{Y_y^2|Q|^2}{LX_x^2}r_2,
\end{equation}
where $t, x, y$ must be expressed in terms of new variables in tildes from \eqref{equiv} by the inverse transformation. We introduce the modulus and the phase of $Q$ in the form $Q=Re^{i\theta}$.
The fact that the coefficients of the system depend only on a single variable $t$ implies that $X$ and $Y$ must be linear in $x$ and $y$, respectively: $X=\alpha(t)x+\alpha_0(t)$, $Y=\beta(t)y+\beta_0(t)$ and also $|Q|=R(t)$, $L=L(t)$.  Using the first two equations of \eqref{constr123} we find
$$2p_1\alpha \theta_x+\dot{\alpha}x+{\alpha}_0=0, \quad 2p_2\beta \theta_y+\dot{\beta}y+{\beta}_0=0,$$
from which we obtain
\begin{equation}\label{teta}
    \theta(t,x,y)=-\frac{\dot{\alpha}}{4p_1\alpha}x^2-\frac{\dot{\beta}}{4p_2\beta}y^2-\frac{\dot{\alpha}_0}{2p_1\alpha}x
    -\frac{\dot{\beta}_0}{2p_2\beta}y+\theta_0(t),
\end{equation}
where $\theta_0(t)$ is an arbitrary integration function. The third expression of \eqref{constr123} gives
$R(t)=\kappa (\alpha\beta)^{1/2}$, $\kappa\ne 0$, $\alpha\beta>0$ (due to the fact that  $H$ is real) and
\begin{equation}\label{H}
    q_2 H=-(\theta_t+p_1\theta_x^2+p_2\theta_y^2).
\end{equation}
Solving this equation for $H$ and putting in the last constraint of \eqref{constr123} and taking into account the vanishing derivatives $\theta_{xy}=\theta_{xyy}=\theta_{yyy}=\theta_{xxx}=\theta_{xxy}=0$ we get the constraint
\begin{equation}\label{constr0}
    2p_1\theta_{xx}^2+2p_2 r_1  \theta_{yy}^2+\theta_{xxt}+r_1 \theta_{yyt}=0.
\end{equation}
To summarize, the equivalence transformations consist of transformations
\begin{equation}\label{finalequiv}
\begin{split}
&\tilde{t}=T(t), \quad \tilde{x}=\alpha(t)x+\alpha_0(t), \quad \tilde{y}=\beta(t)y+\beta_0(t), \\[.5mm]
&    \psi= (\alpha\beta)^{1/2}\exp[i\theta(t,x,y)]\tilde{\psi},  \quad w=L(t)\tilde{w}+H(t,x,y),
\end{split}
\end{equation}
with $\theta$ given by \eqref{teta}, $H$ by \eqref{H} and the constraint \eqref{constr0} satisfied. The set of equivalence transformations form a group. We call it equivalence group and denote  by $\mathsf{G_E}$.

Equivalence transformations mapping \eqref{VCDS} to a constant coefficient DS system is obtained by setting the new coefficients written in tildes in \eqref{newcoeff} to constants.

\subsection{Transformation to canonical form}
We can choose the free functions $\alpha, \beta, L$ in \eqref{finalequiv} as
\begin{equation}\label{alfa-beta-L}
    \alpha=\left(\varepsilon\frac{\dot{T}}{p_1}\right)^{1/2}, \quad \beta=\left(\varepsilon\frac{\dot{T}}{p_2}\right)^{1/2}, \quad L=\frac{\dot{T}}{q_2}, \quad \varepsilon=\pm 1
\end{equation}
to normalize $(p_1,p_2,q_2)$ into $(\tilde{p}_1,\tilde{p}_2,\tilde{q}_2)=(1,1,1)$ provided that $p_1$ and $p_2$ have the same sign. The remaining coefficients transform to
\begin{equation}\label{rest}
    (q_1,r_1,r_2)\to (\tilde{q}_1,\tilde{r}_1,\tilde{r}_2)=\left(\frac{q_1}{\sqrt{p_1p_2}},\frac{p_1}{p_2}r_1,\frac{q_2p_1^{1/2}}{p_2^{3/2}}r_2\right).
\end{equation}
With this choice, the constraint \eqref{constr0} becomes
\begin{equation}\label{cond}
    2 (p_2+p_1 r_1)\{T,t\}+3\frac{p_2 \dot{p}_1^2}{p_1^2}-2\frac{p_2 \ddot{p}_1}{p_1}+r_1\left(3\frac{p_1 \dot{p}_2^2}{p_2^2}-2\frac{p_1 \ddot{p}_2}{p_2}\right)=0,
\end{equation}
where $\{T;t\}$ is the Schwarzian derivative of $T$ with respect to $t$ defined by
$$ \{T;t\}=\frac{\dddot{T}}{\dot{T}}-\frac{3}{2}\Bigl ( \frac{\ddot{T}}{\dot{T}}\Bigr)^2.$$
There are two cases:

I.  $p_2+p_1 r_1=0$ and $T$ is arbitrary: The rest of the constraint \eqref{cond}  gives
\begin{equation}\label{constr}
   2r_1 \dot{p}_1\dot{r}_1+3p_1\dot{r}_1^2-2p_1r_1\ddot{r}_1=0
\end{equation}
and integrates to the relation $\dot{r}_1^2/p_1^2 r_1^3=\text{const.}$ From \eqref{rest} it follows that $\tilde{r}_1=-1$ and $\tilde{r}_1=-\tilde{q}_2$ ($\tilde{q}_2$ being normalized). This corresponds to the integrable case and the equivalence group  $\mathsf{G_E}$ contains 4 arbitrary functions ($T$, $\alpha_0$, $\beta_0$, $\theta_0$) of time.

II. $p_2+p_1 r_1\ne 0$ and $\{T;t\}=0$ or $T$ is a M\"{o}bius transformation in $t$. In this case $\tilde{r}_1\not=-1$ (non-integrable case). The equivalence group  $\mathsf{G_E}$ contains 3 arbitrary functions ($\alpha_0$, $\beta_0$, $\theta_0$) of time. Then condition \eqref{cond} reduces to the differential constraint  between $p_1$, $p_2$ and $r_1$
\begin{equation}\label{constr2}
    p_1^3r_1(3\dot{p}_2^2-2p_2\ddot{p}_2)+p_2^3(3\dot{p}_1^2-2p_1\ddot{p}_1)=0.
\end{equation}

In order to find the conditions on the coefficients that map the system to a constant coefficient or standard DS equations we set all the transformed coefficients $(\tilde{p}_1, \tilde{p}_2, \ldots)\to \text{constants}$  and eliminate the arbitrary functions $T$, $\alpha$, $\beta$, and $L$. This gives us the conditions
\begin{equation}\label{sDS-conds}
    \frac{p_1 p_2}{q_1^2}=A=\text{const.}, \quad \frac{p_1 q_2^2 r_2^2}{p_2^3}=B=\text{const.}
\end{equation}
in addition to
\begin{equation}\label{sDS-conds-2}
  \dot{r}_1^2/(p_1^2 r_1^3)=\text{const.}, \quad  p_2+p_1 r_1=0
\end{equation}
as in case I. These relations also imply $\dot{r}_1/(q_1 r_1)=\text{const.}$ and $(q_1q_2r_2)/p_2^2=\text{const.}$ The relevant equivalence transformation will take \eqref{VCDS} to the one with the coefficients
\begin{equation}\label{const-coeff}
  (\tilde{p}_1, \tilde{p}_2, \tilde{q}_1, \tilde{q}_2, \tilde{r}_1, \tilde{r}_2)=(1,1,A,1,-1,B).
\end{equation}

In case II only the conditions \eqref{sDS-conds} and \eqref{constr2} must be satisfied. $r_1$ cannot be transformed to a constant unless the expression  $(p_1 r_1)/p_2$ is a constant equal to $-1$.

\subsection{The symmetry group of the DS equations}
As a byproduct of the equivalence transformations of the vcDS equations we can obtain the full Lie symmetry group of the DS equations. Exponentiation of the Lie symmetry algebra of the system is often used  to obtain the symmetry group in the literature. However, for systems in more than two spacial dimensions where the symmetry algebra turns out to be infinite-dimensional and depends on several arbitrary functions, this approach requires integrating elements of the algebra involving arbitrary functions which can be very tricky. This was indeed the case for the DS system and KP equation among others. Indeed, the authors of \cite{ChampagneWinternitz1988} made extra effort to put the symmetry transformations into a more manageable form by introducing new arbitrary functions.

Here we take a different approach to generate the symmetry transformations. We simply put
$$\tilde{p}_1=p_1=1, \quad  \tilde{p}_2=p_2=\varepsilon_1, \quad \tilde{q}_1=q_1=\varepsilon_2, \quad \tilde{q}_2=q_1=1, \quad \tilde{r}_1=r_1=\delta_1, \quad \tilde{r}_2=r_2=\delta_2$$ and solve the equations \eqref{newcoeff} for the arbitrary functions in the form
$$\alpha(t)=\beta(t)=K(t)=\sqrt{\dot{T}}, \quad L(t)=\dot{T}.$$ Note that this is the constant coefficient case and both the condition \eqref{constr} and \eqref{constr2} are satisfied. We demand the system to be invariant under  arbitrary time reparametrizations    which occur only under the condition $p_2+p_1r_1=\varepsilon_1+\delta_1=0$ (this condition actually   ensures that  the DS system is integrable).
In this case we find that the symmetry group depends on four arbitrary functions $T, \alpha_0, \beta_0, \theta_0$ (case I) having the form
\begin{subequations}
\begin{equation}\label{symmgroup}
\begin{split}
\tilde{t}&=T(t), \quad \tilde{x}=\sqrt{\dot{T}}x+\alpha_0(t), \quad \tilde{y}=\sqrt{\dot{T}}y+\beta_0(t), \\
\psi&=\sqrt{\dot{T}}\exp[i\theta(t,x,y)]\tilde{\psi}(\tilde{t},\tilde{x},\tilde{y}),  \quad w=\dot{T}\tilde{w}(\tilde{t},\tilde{x},\tilde{y})+H(t,x,y),
\end{split}
\end{equation}
where
\begin{equation}\label{teta-H}
\begin{split}
\theta(t,x,y)&=-\frac{1}{2\sqrt{\dot{T}}}(\dot{\alpha}_0x+\varepsilon_1 \dot{\beta}_0y)-\frac{\ddot{T}}{8\dot{T}}(x^2+\varepsilon_1y^2)+\theta_0(t),\\
H(t,x,y)&=\frac{1}{8} \{T;t\} (x^2+\varepsilon_1y^2)-\frac{\ddot{T}}{2\dot{T}^{3/2}}(\dot{\alpha}_0 x+\dot{\beta}_0 y)+\frac{\ddot{\alpha}_0x}{2\sqrt{\dot{T}}}+\frac{\ddot{\beta}_0y}{2\varepsilon_1\sqrt{\dot{T}}}-\\
&-\frac{\dot{\alpha}_0^2}{4\dot{T}}-\frac{\dot{\beta}_0^2}{4\varepsilon_1\dot{T}}-\theta_0.
\end{split}
\end{equation}
\end{subequations}
These transformations have the property that whenever the pair $(\tilde{\psi} ,\tilde{w})$ satisfies the DS equations, so does the pair $(\psi, w)$.

We remark that if  we have $\varepsilon_1+\delta_1\ne 0$ (non-integrable case) then we have to weaken the symmetry group to a subgroup by restricting the transformation of $T$ to M\"{o}bius one so that $\{T;t\}=0$. In this case we replace $T$ and its derivatives using the formulas
$$T=\frac{at+b}{ct+d}, \quad \dot{T}=(ct+d)^{-2}, \quad \frac{\ddot{T}}{\dot{T}}={-2c}{(ct+d)^{-1}}, \quad ad-bc=1.$$
In addition, choosing the arbitrary  functions $\alpha_0=\beta_0=\theta_0=0$ (hence $H$ vanishes identically) we recover the $\SL(2,\mathbb{R})$ group action on the solutions of the DS system
\begin{equation}\label{sl2}
\begin{split}
& \psi=(ct+d)^{-1}\exp\left[\frac{ic(x^2+\varepsilon_1 y^2)}{4(ct+d)}\right]\tilde{\psi}\left(\frac{at+b}{ct+d},\frac{x}{ct+d},\frac{y}{ct+d}\right),\\[.5mm]
&w=(ct+d)^{-2}\tilde{w}\left(\frac{at+b}{ct+d},\frac{x}{ct+d},\frac{y}{ct+d}\right),
\end{split}
\end{equation}
where $a$, $b$, $c$ and $d$ are real group parameters with the property $ad-bc=1$. Note that this formula holds both integrable and nonintegrable cases. This special invariance (the so-called pseudo-conformal) was used for constructing exact blow-up solutions for the DS \cite{Ozawa1992, CipolattiKavi2001} and the nonlinear Schr\"{o}dinger equations \cite{CazenaveWeissler1991, KavianWeissler1994}.

Putting $T=t$, $\alpha_0=\beta_0=0$ we obtain the gauge symmetry group  (arbitrary time dependent phase change in $\psi$ and a shift in $w$)
$$\psi(t,x,y)=e^{i\theta_0(t)}\tilde{\psi}(t,x,y), \quad w(t,x,y)=\tilde{w}(t,x,y)-\dot{\theta}_0(t).$$

We note that a direct approach which does not rely on use of Lie symmetry algebra to the derivation of transformations \eqref{symmgroup} and \eqref{sl2} was applied for a 3-wave component version of the DS system in \cite{LiYeChen2008}.

In order to illustrate  the utility of  the above-mentioned process  the symmetry group of the KP (Kadomtsev--Petviashvili) equation will be re-derived. We first construct the equivalence group of the variable coefficient KP equation
\begin{equation}\label{VCKP}
    (u_t+p(t)uu_x+q(t)u_{xxx})_x+\sigma(t)u_{yy}=0.
\end{equation}
We use the results of Ref. \cite{GungorWinternitz2002} and write the equivalence group in the form
\begin{equation}\label{KPequiv}
\begin{array}{ll}
& u(x,y,t)=R(t)\tilde{u}(\tilde{x},\tilde{y},\tilde{t})-
\displaystyle\frac{\dot{\alpha}}{\alpha p}x+S(y,t),\\[.3cm]
& \tilde{t}=T(t),  \quad \tilde{x}=\alpha(t)x+\beta(y,t),\quad \tilde{y}=\nu(t)y+\delta(t),
\\[.3cm]
& \alpha\ne 0,\quad  R\ne 0\quad \nu\ne 0,\quad \dot{T}\ne
0.
\end{array}
\end{equation}
The coefficients in the transformed equation satisfy
\begin{equation}\label{newcoefKP}
  \tilde p(\tilde t) = p(t)\frac{{R\alpha }}
{{\dot T}},\quad \tilde q(\tilde t) = q(t)\frac{{\alpha ^3 }}
{{\dot T}}, \quad
  \tilde \sigma (\tilde{t})   = \sigma (t)\frac{{\nu^2 }}
{{\alpha \dot T}}.
\end{equation}
Moreover, the coefficients and the functions in the transformation \eqref{KPequiv} must satisfy some determining equations which we do not reproduce here.
Now we set $\tilde{p}=p=1$, $\tilde{q}=q=1$, $\tilde{\sigma}=\sigma=\sigma_0=\text{const.}$ and solve \eqref{newcoefKP} and the determining equations. This leads us to the symmetry group of the KP equation (in fact self-equivalence)
$$(u_t+uu_x+u_{xxx})_x+\sigma_0u_{yy}=0.$$
The symmetry group depends on three arbitrary functions $T(t)$, $\delta(t)$, $\mu(t)$ and can be expressed as follows:
\begin{equation}\label{SymmGroupKP}
\begin{split}
   \tilde{t}&=T(t),  \quad \tilde{x}=\dot{T}^{1/3}x-\frac{\ddot{T}y^2}{6\sigma_0\dot{T}^{2/3}}-\frac{\dot{\delta}y}{2\sigma_0\dot{T}^{1/3}}+\mu(t),\quad \tilde{y}=\dot{T}^{2/3}y+\delta(t),\\
 u&=\dot{T}^{2/3}\tilde{u}(\tilde{x},\tilde{y},\tilde{t})+\frac{1}{6\sigma_0} \left(\frac{\dddot{T}}{\dot{T}}-\frac{4}{3}\frac{\ddot{T}^2}{\dot{T}^2}\right)y^2+\frac{1}{2\sigma_0}\left(\frac{\ddot{\delta}}{\dot{T}^{2/3}}-
\frac{\dot{\delta}\ddot{T}}{\dot{T}^{5/3}}\right)y\\
& \quad -\frac{\ddot{T}x}{3\dot{T}}-\frac{\dot{\mu}}{3\dot{T}^{1/3}}-
\frac{\dot{\delta}^2}{4\sigma_0\dot{T}^{4/3}}.
\end{split}
\end{equation}
This formula again transforms solutions of the KP (Kadomtsev-Petviashvili) equation among each other and can be used to obtain new solution $\tilde{u}$ involving three arbitrary functions of $t$  from the trivial solution $u$ or vice versa.

In particular, restricting $T(t)$ to M\"{o}bius transformation and putting $\delta(t)=\mu(t)=0$ we obtain the $SL(2,\mathbb{R})$ induced action on the solutions
\begin{equation}\label{sl2-KP}
\begin{split}
\tilde{t}&=\frac{at+b}{ct+d},  \quad ad-bc=1, \quad \tilde{x}=(ct+d)^{-2/3}\left(x+\frac{cy^2}{3\sigma_0(ct+d)}\right),\quad \tilde{y}=(ct+d)^{-4/3}y,\\[.5mm]
 u&= (ct+d)^{-4/3}\tilde{u}(\tilde{x},\tilde{y},\tilde{t})+\frac{c^2y^2}{9\sigma_0(ct+d)^2}+\frac{2cx}{3(ct+d)}.
\end{split}
\end{equation}
We remark that the symmetry group transformations of the KP equation was first obtained in a different form (involving definite integrals of arbitrary functions) in \cite{DavidKamranLeviWinternitz1986} based on integrating the Lie algebra depending on whether $\tau$ is zero or nonzero (see Section \ref{section3}).

\section{Lie point symmetries}\label{section3}
We shall use the usual Lie algorithm for determining the symmetry algebra $L$ of the variable coefficient DS system \eqref{VCDS}.
It is convenient to set  $\psi=u+iv$ and convert \eqref{VCDS} to the following real system of equations:
\begin{subequations}\label{triple}
\begin{eqnarray}
   \label{triple1}  &&\Delta_1\equiv u_t+p_1v_{xx}+p_2v_{yy}=q_1v(u^2+v^2)+q_2vw=0, \\
   \label{triple2}  &&\Delta_2\equiv -v_t+p_1u_{xx}+p_2u_{yy}=q_1u(u^2+v^2)+q_2uw=0,\\
   \label{triple3}  &&\Delta_3\equiv w_{xx}+r_1w_{yy}=r_2(u^2+v^2)_{yy}=0.
\end{eqnarray}
\end{subequations}
A vector field of the form
\begin{equation}
V=\tau\dt+\xi\dx+\eta\dy+\phi_1\du+\phi_2\dv+\phi_3\dw,
\end{equation}
on the jet space $\mathsf J^2(t,x,y,u,v,w)$  will be an element of $L$ if its second order prolongation $\mathsf {Pr}^{(2)}V$ annihilates the system \eqref{triple} on its solution surface:
\begin{equation}\label{inf-inv}
  \mathsf {Pr}^{(2)}V(\Delta_i)\big\vert_{\Delta_j=0}=0,\quad i,j=1,2,3.
\end{equation}
Applying \eqref{inf-inv} with aid of the Mathematica program provides us with an overdetermined  system of linear partial differential equations for $\tau$, $\xi$, $\eta$ and $\phi_i$, $i=1,2,3$  (determining system). Solving the determining system (we skip intervening calculations)
we find that the coefficients of the vector field $V$ must have the form
\begin{equation}
\begin{split}
&\tau=\tau(t),\\
&\xi=\frac{1}{2}\bigl(\dot\tau+\frac{\dot p_1}{p_1}\tau\bigr)x+g(t),\\
&\eta=\frac{1}{2}\Big[\dot\tau+\left(\frac{\dot p_1}{p_1}+\frac{\dot r_1}{r_1}\right)\tau\Big]y+h(t),\\
&\phi_1=-\frac{1}{2}\left(\dot\tau+\frac{\dot q_1}{q_1}\tau\right) u+\Phi v,\\
&\phi_2=-\Phi u-\frac{1}{2}\left(\dot\tau+\frac{\dot q_1}{q_1}\tau\right)v,\\
&\phi_3=-\Big[\dot\tau+\tau \frac{d}{dt}\ln\left(\frac{q_1r_1}{r_2}\right)\Big]w+\frac{1}{q_2}\Phi_t,
\end{split}
\end{equation}
where
\begin{align*}
&\Phi(t,x,y)=-\frac{x^2}{8p_1}\der \Big[\dot\tau+\frac{\dot p_1}{p_1}\tau\Big] -\frac{y^2}{8p_2}\der \Big[\dot\tau+\left(\frac{\dot p_1}{p_1}+\frac{\dot r_1}{r_1}\right)\tau\Big]        -\frac{\dot g}{2p_1}x-\frac{\dot h}{2p_2}y+m(t)
\end{align*}
and $g(t)$, $h(t)$ and $m(t)$ are arbitrary smooth functions of $t$.
The coefficients  $p_1,p_2,q_1,q_2,r_1,r_2$  and the function $\tau(t)$ must satisfy the remaining determining system
\begin{subequations}\label{remaining}
\begin{align}
&\tau\big(\frac{\dot p_1}{p_1}-\frac{\dot p_2}{p_2}+\frac{\dot r_1}{r_1}\big)=0,\label{det1} \\
&\tau\big(\frac{\dot q_1}{q_1}-\frac{\dot q_2}{q_2}+\frac{\dot r_1}{r_1}-\frac{\dot r_2}{r_2}\big)=0,\label{det2}\\
&\frac{d}{dt}\Big[\tau\big(\frac{\dot p_1}{p_1}-\frac{\dot q_1}{q_1}+\frac{\dot r_1}{2r_1}\big)\Big]=0,\label{det3}\\
&A_0\tau+A_1\dot\tau+A_2\ddot\tau+A_3\dddot\tau=0,  \label{det4}
\end{align}
\end{subequations}
where
\begin{subequations}\label{A}
\begin{align}
A_0&=\frac{2\dot p_1^3}{p_1^3}+\frac{3 p_2 \dot p_1^3}{p_1^4r_1}+\frac{\dot p_1^2\dot p_2}{p_1^2p_2}+\frac{\dot p_2 \dot r_1^2}{p_2r_1^2}
    +\frac{2\dot r_1^3}{r_1^3}-\frac{3\dot p_1 \ddot p_1}{p_1^2}-\frac{4p_2\dot p_1 \ddot p_1}{p_1^3r_1}-\frac{\ddot p_1 \dot p_2 }{p_1 p_2}\nonumber\\
    &-\frac{\ddot r_1 \dot p_2 }{r_1 p_2}-\frac{3\dot r_1\ddot r_1}{r_1^2}+\frac{\dddot p_1}{p_1}+\frac{p_2 \dddot p_1}{p_1^2 r_1}
    +\frac{\dddot r_1}{r_1},\label{A0}\\
A_1&=-\frac{2\dot p_1^2}{p_1^2}-\frac{3p_2 \dot p_1^2}{p_1^3r_1}-\frac{\dot p_1 \dot p_2}{p_1 p_2}
     -\frac{ \dot r_1 \dot p_2 }{ r_1  p_2} - \frac{2\dot r_1^2}{r_1^2} + \frac{2\ddot p_1}{p_1} + \frac{2 p_2 \ddot p_1}{ p_1^2r_1}
     +\frac{2\ddot r_1}{r_1},\label{A1}\\
A_2&=\frac{\dot p_1}{p_1}-\frac{\dot p_2}{p_2}+\frac{\dot r_1}{r_1},\label{A2}\\
A_3&=1+\frac{p_2}{p_1r_1}.\label{A3}
\end{align}
\end{subequations}
We shall analyze the system of ODEs \eqref{remaining} for consistent solutions.

There are two subcases.
\par{(a)}  The case $\tau(t)\ne 0$: Let $\tau$ remain free. We are then expected to detect the integrable subclasses.   From \eqref{det4} it follows that we  must have $A_0=A_1=A_2=A_3=0$. From $A_2=A_3=0$ we have $p_2=-p_1r_1$. We observe that under the condition $A_2=0$ or $A_3=0$  we  find that $A_0$ is a differential consequence of $A_1$: $A_0=\dot{A}_1/2$. If  $A_1=0$, then  $A_0=0$ is satisfied automatically. This means that $A_0=0$ is redundant. The remaining three determining equations of \eqref{remaining} are
\begin{equation}\label{rest-deteqs}
    \frac{\dot p_1}{p_1}-\frac{\dot p_2}{p_2}+\frac{\dot r_1}{r_1}=0, \quad \frac{\dot q_1}{q_1}-\frac{\dot q_2}{q_2}+\frac{\dot r_1}{r_1}-\frac{\dot r_2}{r_2}=0, \quad \frac{\dot p_1}{p_1}-\frac{\dot q_1}{q_1}+\frac{\dot r_1}{2r_1}=0.
\end{equation}

We eliminate $\dot{r}_1$ between $A_1=0$ and $A_2=0$ using $p_2=-p_1r_1$ and find again \eqref{constr}
\begin{equation}\label{constr-a}
    2\frac{\ddot{r}_1}{\dot{r}_1}-3\frac{\dot{r}_1}{\dot{r}_1}-2\frac{\dot{p}_1}{\dot{p}_1}=0.
\end{equation}

From $A_2=0$ and the first relation of \eqref{rest-deteqs} we find
\begin{equation}\label{constr-b}
  \frac{\dot p_1}{p_1}+\frac{\dot p_2}{p_2}-2\frac{\dot q_1}{q_1}=0.
\end{equation}
Also from the rest of \eqref{rest-deteqs} using \eqref{constr-b} we find
\begin{equation}\label{constr-c}
  \frac{\dot{q}_1}{q_1}+\frac{\dot{q}_2}{q_2}+\frac{\dot{r}_2}{r_2}-2\frac{\dot{p}_2}{p_2}=0.
\end{equation}
Integrating equations \eqref{constr-a}-\eqref{constr-b} with $p_2=-p_1r_1$ leads to the relation $\dot{r}_1/(q_1r_1)=\text{const.}$ and equation  \eqref{constr-c} to the relation $(q_1q_2r_2)/p_2^2=\text{const.}$ (compare with \eqref{sDS-conds}-\eqref{sDS-conds-2}). This gives us the classification of  the coefficients required for the DS equations to admit a KMV loop algebra (see Tables \ref{tab1}-\ref{tab2}). The existence of four constraints between the coefficients means that two of six arbitrary coefficients can be allowed to remain free (for example $q_1$ and $q_2$).  The four relations obtained are exactly the same for the equation to be equivalent to the integrable DS equation by a local point transformation.

The main consequence of this observation is that we can pick out integrable VCDS equations from a general class of nonintegrable ones.

We give an example of an integrable variable coefficient DS equation with the coefficients defined by
$$p_1=e^t, \quad p_2=e^{-t}, \quad q_1=1, \quad q_2=e^{-4t}, \quad r_1=-e^{-2t}, \quad r_2=e^{2t}.$$
The corresponding equivalence transformation taking this equation to its constant coefficient form is not unique  and can be  obtained from \eqref{finalequiv} choosing the free functions as $T=t$, $\alpha_0=\beta_0=\theta_0=0$:
\begin{equation}
\begin{split}
&\tilde{t}=t, \quad \tilde{x}=e^{-t/2}x, \quad \tilde{y}=e^{t/2}y, \\
&    \psi=\exp[\frac{i}{8}(e^{-t}x^2-e^ty^2)]\tilde{\psi},  \quad w=e^{4t}\tilde{w}+\frac{e^{4t}}{16}(e^{-t}x^2+e^ty^2).
\end{split}
\end{equation}
The coefficients of the transformed equations will be
$$(\tilde{p}_1, \tilde{p}_2, \tilde{q}_1, \tilde{q}_2, \tilde{r}_1, \tilde{r}_2)=(1,1,1,1,-1,1)$$
as can be read off from \eqref{const-coeff}.

\begin{table}
\caption {Classification results in the integrable case. $k_0,k_1,r_{10},r_{20}$ are constants ($r_{10},r_{20}$ have been rescaled).} \label{tab1}
\begin{center}
    \begin{tabular}{| l | l | l | }
    \hline
    $\displaystyle p_1(t)=k_1 q_1(t) e^{k_0\int q_1(t)dt}$ & $\displaystyle q_1(t)=\text{free}$ & $\displaystyle r_1(t)=r_{10} e^{-2k_0\int q_1(t)dt}$ \\ \hline
    $\displaystyle p_2(t)=-r_{10}k_1 q_1(t) e^{-k_0\int q_1(t)dt}$ & $\displaystyle q_2(t)=\text{free}$ & $\displaystyle r_2(t)=r_{20}\frac{q_1(t)}{q_2(t)}e^{-2k_0\int q_1(t)dt}$  \\ \hline
        \end{tabular}
\end{center}
\end{table}

There is another classification of the coefficients if $p_1$ and $q_2$ are assumed to be free.
\begin{table}
\caption {Alternative classification results in the integrable case} \label{tab2}
\begin{center}
    \begin{tabular}{| l | l | l | }
    \hline
    $ p_1(t)=\text{free}$   & $\displaystyle q_1(t)=\frac{p_1(t)}{k_0\int p_1(t)dt+k_1}$   & $\displaystyle r_1(t)=\frac{r_{10}}{(k_0\int p_1(t)dt+k_1)^2}$ \\ \hline
    $\displaystyle p_2(t)=-r_{10}\frac{p_1(t)}{(k_0\int p_1(t)dt+k_1)^2}$ & $q_2(t)=\text{free}$ & $\displaystyle r_2(t)=\frac{r_{20}}{(k_0\int p_1(t)dt+k_1)^3} \frac{p_1(t)}{q_2(t)}$  \\ \hline
        \end{tabular}
\end{center}
\end{table}
In this setting the infinitesimal generator can be written as
\begin{subequations}
\begin{equation}
V=X(\tau)+Y(g)+Z(h)+W(m),
\end{equation}
where
\begin{align}
X(\tau)&=\tau(t)\Big[\dt+\frac{1}{2}(\frac{\dot q_1}{q_1}+k_0 q_1)x\,\dx+\frac{1}{2}(\frac{\dot q_1}{q_1}-k_0 q_1)y\,\dy
                    -\frac{\dot q_1}{2q_1}u\du-\frac{\dot q_1}{2q_1}v\dv-\frac{\dot q_2}{q_2}w\dw\Big]  \nonumber\\
       &+ \frac{\dot\tau(t)}{2}\Big[x\dx+y\dy-u\du-v\dv-2w\dw\Big]+\Phi(v\du-u\dv)+\frac{1}{q_2(t)}\Phi_t\dw\\
Y(g)&=g(t) \dx-\frac{x\dot g (t)}{2p_1(t)}(v \du-u \dv)-\frac{x}{2q_2(t)} \frac{d}{dt}\Big[\frac{\dot{g}(t)}{p_1(t)}\Big]\, \dw,\\
Z(h)&=h(t) \dy - \frac{y \dot h(t)}{2p_2(t)}(v \du-u \dv)-\frac{y}{2q_2}\frac{d}{dt}\Big[\frac{\dot h(t)}{p_2(t)}\Big]\, \dw,\\
W(m)&=m(t)(v\du-u\dv)+\frac{\dot m(t)}{q_2(t)} \,\dw.
\end{align}
\end{subequations}
Here $g(t),h(t)$ and $m(t)$ are arbitrary functions and $\Phi$ is as follows
\begin{equation*}
\Phi(t,x,y)=-\frac{x^2}{8p_1(t)}\der\Big[\dot\tau(t)+\Big(\frac{\dot q_1(t)}{q_1(t)}+k_0 q_1(t)\Big)\tau\Big]-\frac{y^2}{8p_2(t)}\der\Big[\dot\tau(t)+\Big(\frac{\dot q_1(t)}{q_1(t)}-k_0 q_1(t)\Big)\tau(t)\Big].
\end{equation*}
The above generators are formulated for the coefficients given in Table \ref{tab1}.
After some tedious computation we find that the commutation relations among  the generators satisfy
\begin{equation}\label{comm}
\begin{split}
&[X(\tau_1),X(\tau_2)]=X(\tau_1\dot\tau_2-\tau_2\dot\tau_1),\\
&[X(\tau),W(m)]=W(\tau\dot m),\\
&[X(\tau),Y(g)]=Y(G_1),\quad G_1=\tau\dot g-\frac{1}{2}g\dot\tau -\frac{1}{2}g\tau(k_0  q_1+\frac{\dot q_1}{q_1}),\\
&[X(\tau),Z(h)]=Z(G_2),\quad G_2=\tau\dot h-\frac{1}{2}h\dot\tau +\frac{1}{2}h\tau(k_0  q_1-\frac{\dot q_1}{q_1}),\\
&[Y(g_1),Y(g_2)]=W(G_3),\quad G_3=-\frac{e^{-k_0\int q_1 dt}}{2k_1q_1}(g_1\dot g_2-g_2\dot g_1),\\
&[Z(h_1),Z(h_2)]=W(G_4),\quad G_4= \frac{e^{k_0\int q_1 dt}}{2k_1r_{10}q_1}(h_1\dot h_2-h_2\dot h_1),\\
&[Y(g),Z(h)]=[Y(g),W(m)]=[Z(h),W(m)]=[W(m_1),W(m_2)]=0.
\end{split}
\end{equation}
They imply that the integrable vcDS symmetry algebra $L$  admits a  Levi-decomposition
\begin{equation}\label{KMV}
L=\curl{X(\tau)}\rhd \curl{Y(g), Z(h), W(m)},
\end{equation}
where $\curl{X(\tau)}$ is an infinite-dimensional simple algebra and $ \curl{Y(g), Z(h), W(m)}$ is the nilradical of $L$.

In the special case where all coefficients are constants (the standard DS system)
\begin{equation}
p_1(t)=\alpha_1,\quad p_2(t)=\alpha_2, \quad q_1(t)=\beta_1, \quad q_2(t)=\beta_2, \quad r_1(t)=\gamma_1, \quad r_2(t)=\gamma_2,
\end{equation}
an isomorphic Lie algebra is realized when $\alpha_1\gamma_1+\alpha_2=0$. The corresponding commutation relations can be extracted from \eqref{comm} by choosing $k_0=0$ along with a suitable rearrangement of the constants $k_1,r_{10}$. The functions $G_1, G_2, G_3, G_4$ of \eqref{comm} simplify to
$$G_1=\tau\dot g-\frac{1}{2}g\dot\tau, \quad G_2=\tau\dot h-\frac{1}{2}h\dot\tau, \quad G_3=-\frac{1}{2\alpha_1}(g_1\dot g_2-g_2\dot g_1), \quad G_4= -\frac{1}{2\alpha_2}(h_1\dot h_2-h_2\dot h_1).$$

We summarize our results as theorems.
\begin{thm}\label{t1}
The variable coefficient DS (vcDS)  system   \eqref{VCDS} admits the (centerless) Virasoro algebra as
a symmetry algebra if and only if the coefficients satisfy the relations of Table \ref{tab1} or Table \ref{tab2}.
\end{thm}

\begin{thm}\label{t2}
Under the conditions of Table \ref{tab1} or Table \ref{tab2} the symmetry algebra of the vcDS system is isomorphic to that of the  integrable standard DS (sDS) system and a point transformation, a special case of the equivalence transformation \eqref{finalequiv}, exists transforming among each other.
\end{thm}

\par{(b)} We now let $A_3=p_2+p_1 r_1\ne 0$ but $\dddot{\tau}=0$ ($\tau=\tau_2 t^2+\tau_1 t+\tau_0$, $\tau_2, \tau_1, \tau_0$ not all zero). Then the Virasoro part of the symmetry algebra is reduced to the finite-dimensional algebra $\Sl(2,\mathbb{R})$.
In this case, instead of \eqref{constr} (or \eqref{constr-a}) we recover the  relation \eqref{constr2} which results from eliminating  the derivatives of $r_1$ between the relations $A_1=0$, $A_2=0$ (we discard $A_0$). In summary, the relations between the coefficients turn out to be less restrictive (only three relations). The original system can no longer be integrable.

\par  The case $\tau(t)=0$: In this case the  symmetry algebra is generated by
\begin{equation}
V=Y(g)+Z(h)+W(m)
\end{equation}
without any restriction on the coefficients $p_1,p_2,q_1,q_2,r_1,r_2$.

\begin{rmk}
If we apply our results for  the KMV invariance requirements to the canonical vcDS (cvcDS) equation in which $p_1=p_2=q_2=1$, and the other coefficients are arbitrary we find that cvcDS equation  must be  constant coefficient one.

\end{rmk}

\section{Solutions of the vcDS system}\label{section4}

\subsection{Solutions from simple solutions}
One can make use of the symmetry group \eqref{symmgroup} to generate new solutions of the vcDS system from simple ones, for example stationary solutions, regardless of integrability. We looked for lump type stationary solutions which were first constructed and used by Ozawa in \cite{Ozawa1992} to prove the existence of blow-up solutions to the focusing  DS II system (related to \eqref{DS} by the exchange $x \leftrightarrow y$ and $w \leftrightarrow w_y$) based on $\SL(2,\mathbb{R})$ invariance (also called pseudo-conformal invariance).  We found that the vcDS system also admits the lump-type solution
\begin{equation}
\psi(x,y)=\frac{1}{1+\alpha x^2+\beta y^2}, \quad w=\frac{\partial}{\partial y}\left(\frac{2\alpha\gamma y}{1+\alpha x^2+\beta y^2}\right),
\end{equation}
where $\alpha, \beta, \gamma$ are arbitrary constants,  if the coefficients of \eqref{VCDS} satisfy
\begin{equation}
p_2(t)=-\frac{\alpha}{\beta}p_1(t), \quad q_1(t)=-8\alpha p_1(t), \quad q_2(t)=\frac{4}{\gamma}p_1(t), \quad r_1(t)=\frac{\alpha}{\beta}, \quad r_2(t)=\frac{4\alpha^2\gamma}{\beta}.
\end{equation}
These conditions fall into those for the DS to be integrable or transformable to the integrable standard DS. Namely, conditions \eqref{sDS-conds}-\eqref{sDS-conds-2} are satisfied.

The special choice $\alpha=\beta>0$, $p_1>0$ (then  $r_1=1$, $r_2=4\alpha\gamma$ are necessarily constants and $p_2$, $q_1, q_2$ are proportional to an arbitrary $p_1$) corresponds to a focusing ($q_1<0$) integrable case (vcDS I or vcDS II). In this case an application of the symmetry group action \eqref{sl2} to the above-mentioned lump solution $\tilde{\psi}=\Psi_{\alpha}(x,y)=(1+\alpha(x^2+y^2))^{-1}\in L^2(\mathbb{R}^2)$ depending on a parameter $\alpha$  can be used in exactly the same manner as in Ozawa's approach \cite{Ozawa1992} to obtain exact blow up solution $\psi_\alpha\in L^2(\mathbb{R}^2)$, but $\psi_\alpha\not\in H^1$ in finite time for a  Cauchy problem with a special initial data
$\psi(0,x,y)$ for the vcDS system  by an appropriate choice of the group parameters $a,b,c$.

We note that a numerical study of the long time behaviour and potential blow-up solutions to the focusing DS II (hyperbolic-elliptic case) equation has recently appeared in  \cite{KleinMuiteRoidot2011}. Solutions obtained by transforming lump solutions through the symmetry transformations for the vcDS equations can be useful for testing numerical schemes  for such equations.

On the other hand, the equivalence group \eqref{finalequiv} can also be used to transform the stationary solutions  $\psi(x,y)$ and $w(x,y)$  to new solutions of the vcDS system depending on four arbitrary functions of time.
The stationary DS system
\begin{subequations}\label{Stationary}
\begin{eqnarray}
    &&\psi_{xx}+\epsilon_1\psi_{yy}=\epsilon_2|\psi|^2\psi+\psi w, \\
    &&w_{xx}+\delta_1w_{yy}=\delta_2(|\psi|^2)_{yy}
\end{eqnarray}
\end{subequations}
is invariant under a four-dimensional solvable symmetry algebra with structure $L_0=A_{3,3}\oplus A_1$, $A_{3,3}\simeq \mathsf{T}_2\rhd \mathsf{D}$, $\mathsf{D}$ being a dilation generator. A basis for $L_0$ is
$$\curl{ e_1=\dx,\;e_2=\dy,\;e_3=x\dx+y\dy-u\du-v\dv-2w\dw,\;e_4=-v\du+u\dv}.$$
Applying translational symmetry transformations $x\to x-a_1/a_3$ and $y\to y-a_2/a_3$   we can remove $e_1$, $e_2$ from the general symmetry element $V=a_1 e_1+a_2 e_2+a_3 e_3+a_4 e_4$ if $a_1, a_2, a_3\ne 0$. So it is enough to look at reductions by $V=e_3+a e_4$, $a\in \mathbb{R}$. However, the corresponding reduced ODE system (we do not spell out here) is too complicated to be tractable by known standard analytical techniques.

\subsection{Symmetry reduction in the generic case}
We will perform the symmetry reduction in the generic case in which the symmetry algebra is nilpotent ($\tau=0$), namely when  generated by the vector field
\begin{equation}
V=Y(g)+Z(h)+W(m).
\end{equation}
In terms of the invariants $t$ and $\xi=h(t)x-g(t)y$, the reduction formulas (invariant solutions) are given by
\begin{align}
\psi&=\exp\Big\{i\Big[\frac{\dot g}{4p_1 g}\,x^2+\frac{\dot h}{4p_2 h}\,y^2-\frac{m}{g}\,x\Big]\Big\}\Psi(\xi,t), \qquad \xi=h(t)x-g(t)y  \label{red-form-1}\\
w&=-\frac{1}{4q_2g}\frac{d}{dt}\big(\frac{\dot g}{p_1}\big)\,x^2
   -\frac{1}{4q_2h}\frac{d}{dt}\big(\frac{\dot h}{p_2}\big)\,y^2+\frac{\dot m}{q_2 g}\, x + \widehat{W}(\xi,t). \label{red-form-2}
\end{align}
Substitution of \eqref{red-form-1} and \eqref{red-form-2} into \eqref{VCDS}
leads to the reduced system
\begin{align}
i\Psi_t&+(p_1h^2+p_2g^2)\Psi_{\xi\xi}+i\Big[\big(\frac{\dot g}{g}+\frac{\dot h}{h}\big)\xi-\frac{2p_1hm}{g}\Big]\Psi_\xi
        +\Big[-\frac{p_1m^2}{g^2} + \frac{i}{2}\big(\frac{\dot g}{g}+\frac{\dot h}{h}\big)\Big]\Psi \nonumber \\
&= q_1 |\Psi|^2\Psi+q_2 \Psi \widehat{W},   \label{red1}\\[.5mm]
(r_1g^2&+h^2)\widehat{W}_{\xi\xi}=r_2g^2(|\Psi|^2)_{\xi\xi}+\nu(t), \label{red2}
\end{align}
$$\nu(t)\equiv\frac{\ddot g}{2p_1q_2g}+\frac{r_1\ddot h}{2p_2q_2h}-\frac{r_1\dot h \dot p_2}{2p_2^2q_2h}
-\frac{\dot p_1 \dot g}{2 p_1^2 q_2 g}.$$
Equation \eqref{red2} can be integrated twice with respect to $\xi$ to obtain
\begin{equation}\label{W}
\widehat{W}(\xi,t)=\frac{r_2g^2}{r_1g^2+h^2} |\Psi|^2
+\frac{\nu}{2(r_1g^2+h^2)}\xi^2+\nu_1\xi +\nu_0.
\end{equation}
Here $\nu_0(t)$ and $\nu_1(t)$ are arbitrary functions. Substituting \eqref{W}  into  equation \eqref{red1} we obtain a generalized nonlinear Schr\"{o}dinger equation for $\Psi$ with  coefficients depending on $\xi$ and time:
\begin{equation}\label{NLS}
i\Psi_t+(p_1h^2+p_2g^2)\Psi_{\xi\xi}+ik_2(\xi,t)\Psi_\xi+\Big[h_1(\xi,t)+ ih_2(t)\Big]\Psi=(q_1+\frac{q_2r_2g^2}{r_1g^2+h^2}) |\Psi|^2\Psi,
\end{equation}
where
\begin{subequations}
\begin{align}
k_2(\xi,t)&=k_{21}(t)\xi+k_{20}(t)=\big(\frac{\dot g}{g}+\frac{\dot h}{h}\big)\xi-\frac{2p_1hm}{g},\\
h_1(\xi,t)&=H_2(t)\xi^2+H_1(t)\xi+H_0(t) \nonumber\\
           &=-\frac{q_2\nu}{2(r_1g^2+h^2)}\xi^2-q_2\nu_1\xi -\frac{p_1m^2}{g^2}-q_2\nu_0,\\
h_2(t)&=\frac{1}{2}\big(\frac{\dot g}{g}+\frac{\dot h}{h}\big).
\end{align}
\end{subequations}
The equivalence transformation for \eqref{NLS} is given by
\begin{subequations}\label{NLS-equiv}
\begin{align}
\Psi(\xi,t)&=R(t)\exp[i(\theta_2(t)\xi^2+\theta_1(t)\xi+\theta_0(t))]\Omega(X,T),\\
X&=X_1(t)\xi+X_0(t), \quad T=T(t).
\end{align}
\end{subequations}
Transformation \eqref{NLS-equiv} contains enough freedom to transform the coefficients  $\Psi$ and $\Psi_\xi$ to zero by a choice of $X_0, X_1, \theta_0, \theta_1, \theta_2$
as  solution of the following ODEs
\begin{equation}
\begin{split}
&\frac{\dot X_1}{X_1}+4f\theta_2+k_{21}=0,\quad
\frac{\dot X_0}{X_1}+2f\theta_1+k_{20}=0,\quad
\frac{\dot R}{R}+h_2+2f\theta_2=0,\\
&\dot\theta_0=H_0-k_{20}\theta_1-f\theta_1^2,\quad
\dot\theta_1+(k_{21}+4f\theta_2)\theta_1+2k_{20}\theta_2=H_1,\quad
\dot\theta_2+2k_{21}\theta_2+4f\theta_2^2=H_2,
\end{split}
\end{equation}
where we have defined  $f\equiv p_1h^2+p_2g^2$.
Note that $\theta_2$ satisfies a Riccati equation which can be solved only in special cases.
The reduced equation \eqref{NLS}  is hence
simplified to the following form
\begin{equation}
i\Omega_T+(p_1h^2+p_2g^2)\frac{X_1^2}{\dot T}\Omega_{XX}=\left(q_1+\frac{q_2r_2g^2}{r_1g^2+h^2}\right)\frac{R^2}{\dot T} |\Omega|^2\Omega.
\end{equation}
If we furthermore set $\displaystyle X_1^2=\pm \dot Tf^{-1}$, the coefficient of $\Omega_{XX}$ is normalized to unity so that we arrive at the time-dependent  NLS (Nonlinear Schr\"odinger) equation
\begin{equation}
i\Omega_T+\Omega_{XX}=N(t)|\Omega|^2\Omega,
\end{equation}
where $\displaystyle N(t)=\left(q_1+\frac{q_2r_2g^2}{r_1g^2+h^2}\right)\frac{R^2}{\dot T}$.

The above choice of $X_1$ puts a restriction on  $T(t)$
\begin{equation}
\frac{\ddot T}{\dot T}-\frac{\dot f}{f}+8 f \theta_2+2k_{21}=0,
\end{equation}
which can be integrated to yield
\begin{equation}
T(t)=T_1\int f(t) \exp\left[-\int(8f\theta_2+2k_{21})dt\right]dt+T_0.
\end{equation}

\section{Conclusions}\label{section5}
The usual DS equations are known to allow an infinite dimensional symmetry algebra with a Kac-Moody-Virasoro structure when it is completely integrable. In the same spirit,
we have analyzed the conditions on the coefficients of the variable coefficient DS equations when the symmetry algebra preserves the same loop structure.  We  have shown that under these conditions the corresponding system is equivalent to the integrable DS one by a local point transformation. Consequently, this special class of variable coefficient DS equations can be considered as completely integrable variable coefficient generalization. We have reproduced the symmetry group of the DS system by using the equivalence group of the vcDS system without having recourse to integrating the symmetry algebra.

We have discussed how the equivalence group and the symmetry group can be used to generate solutions depending on four arbitrary functions from stationary solutions like lump solutions of Ozawa type. In the generic case we have presented reduction to the NLS equation.

Finally, we would like to emphasize that it would be of great interest to rework the derivation of the DS equations under different variable physical conditions. The individual variable coefficients figuring in the system would then make  physical  sense.


\begin{thebibliography}{10}

\bibitem{DaveyStewartson1974}
A.~Davey and K.~Stewartson.
\newblock On three-dimensional packets of surface waves.
\newblock {\em Proceedings of the Royal Society of London A: Mathematical,
  Physical and Engineering Sciences}, 338(1613):101--110, 1974.

\bibitem{GhidagliaSaut1990}
J.~M. Ghidaglia and J.~C. Saut.
\newblock On the initial value problem for the {Davey--Stewartson} systems.
\newblock {\em Nonlinearity}, 3(2):475, 1990.

\bibitem{ChampagneWinternitz1988}
B.~Champagne and P.~Winternitz.
\newblock On the infinite--dimensional symmetry group of the
  {Davey--Stewartson} equations.
\newblock {\em Journal of Mathematical Physics}, 29(1):1--8, 1988.

\bibitem{Omote1988}
M. Omote.
\newblock Infinite--dimensional symmetry algebras and an infinite number of
  conserved quantities of the (2+1)--dimensional {Davey--Stewartson} equation.
\newblock {\em Journal of Mathematical Physics}, 29(12):2599--2603, 1988.

\bibitem{GanesanLakshmanan1987}
S. ~Ganesan and M. ~Lakshmanan.
\newblock Singularity-structure analysis and {Hirota}'s bilinearisation of the
  {Davey--Stewartson} equation.
\newblock {\em Journal of Physics A: Mathematical and General}, 20(17):L1143,
  1987.

\bibitem{GuengoerAykanat2006}
F.~G\"ung\"or and \"O.~Aykanat.
\newblock The generalized Davey-Stewartson equations, its
  {Kac--Moody--Virasoro} symmetry algebra and relation to {Davey--Stewartson}
  equations.
\newblock {\em Journal of Mathematical Physics}, 47(1):013510, 2006.

\bibitem{LiYeChen2008}
B. Li, W. Ye and Y. Chen.
\newblock Symmetry, full symmetry groups, and some exact solutions to a
  generalized {Davey--Stewartson} system.
\newblock {\em Journal of Mathematical Physics}, 49(10):103503, 2008.

\bibitem{ZhouTianMoLiWang2013}
H. Zhou, B. ~Tian, H. Mo, M. Li and P. Wang.
\newblock Baecklund transformation, {Lax} pair and solitons of the
  (2+1)-dimensional {Davey--Stewartson}-like equations with variable
  coefficients for the electrostatic wave packets.
\newblock {\em Journal of Nonlinear Mathematical Physics}, 20(1):94--105, 2013.

\bibitem{Ozawa1992}
T.~Ozawa.
\newblock Exact blow--up solutions to the {Cauchy} problem for the
  {Davey--Stewartson} system.
\newblock {\em Proc. Roy. Soc. Lond. A}, 436:345--349, 1992.

\bibitem{GungorOzemir2015}
F.~G\"ung\"or and C.~\"Ozemir.
\newblock {L}ie symmetries of a generalized {Kuznetsov--Zabolotskaya--Khokhlov}
  equation.
\newblock {\em J. Math. Anal. and Appl.}, 423(1):623 -- 638, 2015.
\newblock arXiv:1402.1941.

\bibitem{CipolattiKavi2001}
R. Cipolatti and O. Kavian.
\newblock Existence of pseudo-conformally invariant solutions to the
  {Davey--Stewartson} system.
\newblock {\em Journal of Differential Equations}, 176(1):223 -- 247, 2001.

\bibitem{CazenaveWeissler1991}
T. Cazenave and F. ~B. Weissler.
\newblock The structure of solutions to the pseudo-conformally invariant
  nonlinear {Schr\"odinger} equation.
\newblock {\em Proceedings of the Royal Society of Edinburgh, Section: A
  Mathematics}, 117:251--273, 1991.

\bibitem{KavianWeissler1994}
O.~ Kavian and F. ~B. Weissler.
\newblock Self-similar solutions of the pseudo-conformally invariant nonlinear
  {Schr\"odinger} equation.
\newblock {\em Michigan Math. J.}, 41(1):151--173, 1994.

\bibitem{GungorWinternitz2002}
F.~G{\"u}ng{\"o}r and P.~Winternitz.
\newblock {G}eneralized {Kadomtsev--Petviashvili} equation with an infinite
  dimensional symmetry algebra.
\newblock {\em J. Math. Anal. and Appl.}, 276:314--328, 2002.

\bibitem{DavidKamranLeviWinternitz1986}
D.~David, N.~Kamran, D.~Levi and P.~Winternitz.
\newblock Symmetry reduction for the {Kadomtsev--Petviashvili} equation using a
  loop algebra.
\newblock {\em Journal of Mathematical Physics}, 27(5):1225--1237, 1986.

\bibitem{KleinMuiteRoidot2011}
C.~Klein, B.~Muite and K.~Roidot.
\newblock Numerical study of blowup in the {Davey--Stewartson} system.
\newblock arXiv:1112.4043.

\end{thebibliography}

\end{document}